\documentclass[12pt]{article}

\usepackage{amsmath,amssymb,bm,graphicx}
\usepackage{graphicx}
\usepackage[dvips]{color} 

\newcommand{\pslash}{\not \! p}
\newcommand{\kslash}{\not \! k}

\begin{document}

\Large
\begin{center}
{{\bf BCS, Nambu-Jona-Lasinio, and  Han-Nambu~\footnote{An introductory lecture given at Nambu Memorial Symposium at Osaka City University, September 29, 2015.}\\
---- A sketch of Nambu's works in 1960-1965 ----
}}
\end{center}
\vskip .5 truecm
\begin{center}
{{ Kazuo Fujikawa}}
\end{center}

\begin{center}
\vspace*{0.4cm} { {RIKEN Nishina Center, Wako 351-0198, Japan
}}
\end{center}

\begin{abstract}
The years of 1960-1965 were a remarkable period for Yoichiro Nambu.
Starting with a reformulation of BCS theory with emphasis on gauge invariance, he recognized the realization
of spontaneous chiral symmetry breaking in particle physics as is evidenced by the 
Goldberger-Treiman relation. A concrete model of Nambu and Jona-Lasinio illustrated the essence of the Nambu-Goldstone theorem and the idea of soft pions. After the proposal of the quark model by Gell-Mann, he together with Han constructed an alternative model of integrally charged quarks with possible non-Abelian gluons. All those remarkable works were performed during the years 1960-1965.  Here I briefly review those works following  the original papers of Nambu chronologically, together with a brief introduction to a formulation of Neother's theorem and Ward-Takahashi identities using path integrals.
This article is mostly based on a lecture given at the Nambu Memorial Symposium held at Osaka City University in September 2015, where Nambu started his professional career.   
\end{abstract}

\normalsize

\makeatletter
\@addtoreset{equation}{section}
\def\theequation{\thesection.\arabic{equation}}
\makeatother

\section{Introduction}
Y. Nambu started his professional career at Osaka City University in 1949, then a new university. His theory group at that time included S. Hayakawa, Y. Yamaguchi, K. Nishijima and T. Nakano, and all of them made major contributions to particle physics worldwide and physics in general in Japan. He left Japan in early 1950s and went to Institute for Advanced Study at Princeton and then to University of Chicago, where he stayed till the end of his professional career.

Nambu made numerous original contributions to physics, but in this article which is based on an introductory lecture given at Osaka City University at the end of September, 2015, I concentrate on his works during the years 1960 to 1965.
These years were remarkable years in which he reformulated BCS theory of superconductivity and then applied the idea of spontaneous symmetry breaking to the physical vacuum and particle physics. He also initiated the idea of non-Abelian gluons coupled to quarks.

All those works are well-known and most people in particle physics community are familiar with their contents. Nevertheless, following the suggestion of the organizers of Nambu Memorial Symposium, I briefly review those original papers of Nambu chronologically, together with a brief introduction to a formulation of Neother's theorem and Ward-Takahashi identities using path integrals.

\section{BCS theory and gauge invariance}

The starting Hamiltonian of BCS theory is given by~\cite{BCS} 
\begin{eqnarray}
H&=&\int \sum_{i=1}^{2}\psi_{i}^{\dagger}(x)K_{i}\psi_{i}(x)d^{3}x
+\frac{1}{2}\int\int\sum_{i,k}
\psi_{i}^{\dagger}(x)\psi_{k}^{\dagger}(y)V(x,y)\psi_{k}(y)\psi_{i}(x) d^{3}xd^{3}y\nonumber\\
&=&H_{0}+H_{int}.
\end{eqnarray}
where the indices of $\psi_{i}(x)$, $i=1,2$, stand for spin up and down states.

According to Nambu~\cite{nambu-BCS}, the essence of BCS theory is the Hartree-Fock method.
The linearized Hamiltonian in this method is written as 
\begin{eqnarray}
H^{\prime}_{0}&=&\int \sum_{i=1}^{2}\psi_{i}^{\dagger}(x)K_{i}\psi_{i}(x)d^{3}x\nonumber\\
&+&\frac{1}{2}\int\int\sum_{i,k}d^{3}xd^{3}y[
\psi_{i}^{\dagger}(x)\chi_{ik}(x,y)\psi_{k}(y)+\psi_{i}^{\dagger}(x)\phi_{ik}(x,y)\psi_{k}^{\dagger}(y)
+\psi_{k}(x)\phi^{\dagger}_{ki}(x,y)\psi_{i}(y)]\nonumber\\
&=&H_{0}+H_{s}.
\end{eqnarray}
and the self-consistent equation to be satisfied is
\begin{eqnarray}
\chi_{ik}(x,y)&=&\delta_{ik}\delta(x-y)\int V(x,z)\sum_{j}\langle\psi_{j}^{\dagger}(z)\psi_{j}(z)\rangle d^{3}z-V(x,y)
\langle\psi_{k}^{\dagger}(y)\psi_{i}(x)\rangle,\nonumber\\
\phi^{\dagger}_{ik}(x,y)&=&\frac{1}{2}V(x,y)\langle\psi_{k}^{\dagger}(y)\psi_{i}^{\dagger}(x)\rangle,\nonumber\\
\phi_{ik}(x,y)&=&\frac{1}{2}V(x,y)\langle\psi_{k}(y)\psi_{i}(x)\rangle.
\end{eqnarray}
Under the gauge transformation, these variables are transformed as 
\begin{eqnarray}
&&\psi_{i}(x)\rightarrow e^{i\lambda(x)}\psi_{i}(x),\nonumber\\
&&\chi(x,y)\rightarrow e^{-i\lambda(x)+i\lambda(y)}\chi(x,y),\nonumber\\
&&\phi(x,y)\rightarrow e^{i\lambda(x)+i\lambda(y)}\phi(x,y),\nonumber\\
&&\phi^{\dagger}(x,y)\rightarrow e^{-i\lambda(x)-i\lambda(y)}\phi^{\dagger}(x,y).
\end{eqnarray}
The variable $\phi(x,y)$, whose expectation value 
\begin{eqnarray}
\langle\phi(x,y)\rangle \neq 0 
\end{eqnarray}
characterizes the superconductivity,
carries two units of electronic charge and it is rather involved to satisfy the gauge invariance condition.
The analysis of the Meissner effect is not clear either.

Nambu thus looked at the problem from the point of view of the perturbation theory of Feynman and Dyson, and discussed the issue of gauge invariance.
The starting Lagrangean he adopted is 
\begin{eqnarray}
{\cal L}&=&\sum_{p,i}[i\psi^{\dagger}_{i}(p)\dot{\psi}_{i}(p)-\psi^{\dagger}_{i}(p)\epsilon_{p}\psi_{i}(p)]+\sum_{k}\frac{1}{2}[\dot{\varphi}(k)\dot{\varphi}(-k) -c^{2}k^{2}\varphi(k)\varphi(-k)]\nonumber\\
&&-g\frac{1}{\sqrt{V}}\sum_{p,k}\psi_{i}^{\dagger}(p+k)\psi_{i}(p)h(k)\varphi(k)
\end{eqnarray}
with $\varphi(k)$ standing for the phonon field with the energy $\omega_{k}=ck$. This Lagarangean is rewritten by defining 
\begin{eqnarray}
\Psi(x)=\left(\begin{array}{c}
            \psi_{\uparrow}(x)\\
            \psi^{\dagger}_{\downarrow}(x)
            \end{array}\right),\ \  {\rm or} \ \  
\Psi(p)=\left(\begin{array}{c}
            \psi_{\uparrow}(p)\\
            \psi^{\dagger}_{\downarrow}(-p)
            \end{array}\right),            
\end{eqnarray}
as
\begin{eqnarray}
{\cal L}&=&\sum_{p}\Psi^{\dagger}(p)[i\frac{\partial}{\partial t}-\epsilon_{p}\tau_{3}]\Psi(p)+\sum_{k}\frac{1}{2}[\dot{\varphi}(k)\dot{\varphi}(-k) -c^{2}k^{2}\varphi(k)\varphi(-k)]\nonumber\\
&&-g\frac{1}{\sqrt{V}}\sum_{p,k}\Psi^{\dagger}(p+k)\tau_{3}\Psi(p)h(k)\varphi(k)+\sum_{p}\epsilon_{p}
\end{eqnarray}
with Pauli matrices,
\begin{equation}
\tau_{1}=\left(\begin{array}{cc}
            0&1\\
            1&0
            \end{array}\right),\ \
\tau_{2}=\left(\begin{array}{cc}
            0&-i\\
            i&0
            \end{array}\right),\ \
\tau_{3}=\left(\begin{array}{cc}
            1&0\\
            0&-1
            \end{array}\right).
\end{equation}
Since the gauge invariance is preserved if one sums a suitable set of loop diagrams, he analyzed the issue of gauge invariance by a modified Hartree-Fock approximation.
\\

This paper stands for a summary of the deep understanding of BCS theory by Nambu, 
and he fully utilized his past experiences of Nambu-Bethe-Salpeter equation~\cite{nambu-bethe-salpeter} and the theory of plasmons. It is not easy for a non-expert to follow the details of calculations; it would take at least as much time as Nambu spent on this paper.  He thus clarified\\
(i) When the gap is generated by the Hartree-Fock type consideration, which is the case if one chooses the attractive force, the collective mode appears.\\
(ii)The collective mode helps to preserve the gauge invariance and the Meissner effect is understood in a manner similar to the theory of plasmon.\\
(iii)The essence of superconductivity is understood in terms of field theory and not specific to the condensed matter physics.

In short, he later mentioned elsewhere "gauge invariance, the energy gap, and the collective excitations are logically related to each other as was shown by the author."

He concluded this analysis by saying:~\cite{nambu-BCS} \\
We have discussed here formal mathematical structure
of the BCS-Bogoliubov theory. The nature of the
approximation is characterized essentially as the
Hartree-Fock method, and can be given a simple interpretation
in terms of perturbation expansion. In the
presence of external fields, the corresponding approximation
insures, if treated properly, that the gauge
invariance is maintained. It is interesting that the
quasi-particle picture and charge conservation (or
gauge invariance) can be reconciled at all. This is possible
because we are taking account of the "radiative
corrections" to the bare quasi-particles which are not
eigenstates of charge. These corrections manifest themselves
primarily through the existence of collective
excitations.
\\

He thus contributed to a deeper understanding of BCS theory together with experts such as N.N. Bogoliubov and P.W. Anderson.
In particular, the representation (2.7) is known as {\em Nambu representation} and appears in any standard textbook on superconductivity nowadays. In the more familiar form at finite temperature in condensed matter physics, the Lagrangean is written by assuming locality as 
\begin{eqnarray}
{\cal L}
=\Psi^{\dagger}(\vec{x},\tau)
\left(\begin{array}{cc}
            \partial_{\tau}-\frac{\nabla^{2}}{2m}-\mu &\Delta(\vec{x},\tau)\\
            \Delta^{\star}(\vec{x},\tau)&\partial_{\tau}+\frac{\nabla^{2}}{2m}                +\mu 
            \end{array}\right)\Psi(\vec{x},\tau),
\end{eqnarray}
with
\begin{eqnarray}
\Psi(\vec{x},\tau)=\left(\begin{array}{c}
            \psi_{\uparrow}(\vec{x},\tau)\\
            \psi^{\dagger}_{\downarrow}(\vec{x},\tau)
            \end{array}\right)
\end{eqnarray}
and $0\leq \tau\leq \beta$. Based on this experience in superconductivity and the treatment of spontaneously symmetry broken vacua, he proceeded to a more universal analysis of the manifestation of spontaneous symmetry breaking in wider fields in physics.
\\
\\
I here want to add some personal comments:

If one assumes the Ginsburg-Landau type representation by assuming the locality of the model such as,
\begin{eqnarray}
{\cal L}_{int}&=&\sum_{ki}[\psi_{i}^{\dagger}(x)\phi_{ik}(x)\psi_{k}^{\dagger}(x)
+\psi_{k}(x)\phi^{\dagger}_{ki}(x)\psi_{i}(x)]\nonumber\\
&+&\sum_{ki}[(\partial_{\mu}+2ieA_{\mu}(x))\phi^{\dagger}_{ki}(x)][(\partial_{\mu}-2ieA_{\mu}(x))\phi_{ki}(x)] - V(|\phi_{ki}(x)|)
\end{eqnarray}
the analyses of gauge invariance and Meissner effect (i.e., massive photon)
become much easier, but the essence of BCS theory and the isomer effect, for example, become less transparent. It appears that the effective theoretical analysis does not fit to the taste of Nambu.
\\

It is interesting that the mass term of the right-handed neutrinos (and also the interaction Lagrangean of the neutron-antineutron oscillation) have a structure similar to that of BCS theory,
\begin{eqnarray}
{\cal L}_{\nu - mass}&=&
-\frac{i}{2}m_{R}[\nu^{T}(x)C(1+\gamma_{5})\nu(x) - \overline{\nu}(x)(1-\gamma_{5})C\overline{\nu}^{T}(x)].
\end{eqnarray}
although the neutrinos are electronically neutral.

\section{Chiral symmetry}

Motivated by the analysis of parity violation of Lee and Yang~\cite{lee-yang}, the theory of weak interactions advanced rapidly. The V-A theory of Feynman and Gell-Mann~\cite{feynman-gell-mann} and others emphasized the importance of chiral symmetry in particle physics.
Goldberger and Treiman~\cite{goldberger-treiman} derived the famous Goldberger-Treiman relation on the basis of V-A theory of weak interactions. This derivation itself
is not regarded to be solid nowadays, but it motivated two fundamental papers. One of them is the paper by Nambu on the idea of spontaneous breaking of chiral symmetry and massless soft pions~\cite{nambu-chiral}
, and the other is the linear $\sigma$ model of Gell-Mann and Levy which simultaneously suggested the quark mixing (at that time, neutron-$\Lambda$ mixing)~\cite{gell-mann-levy}.

We first recall the conservation law of the chiral current in Appendix, (A.14),
\begin{eqnarray}
\partial_{\mu}\langle\bar{\psi}(x)\gamma^{\mu}\gamma_{5}\psi(x)\rangle=\langle2mi\bar{\psi}(x)\gamma_{5}\psi(x)\rangle.
\end{eqnarray}
The direct derivation of this relation is to write the left-hand side as
\begin{eqnarray}
&&[(\partial_{\mu}\bar{\psi}(x))\gamma^{\mu}-mi]\gamma_{5}\psi(x)-\bar{\psi}(x)\gamma_{5}[\gamma^{\mu}\partial_{\mu}+im]\psi(x)+2mi\bar{\psi}(x)\gamma_{5}\psi(x)\nonumber\\
&&=2mi\bar{\psi}(x)\gamma_{5}\psi(x),
\end{eqnarray}
ans use the equations of motion $[(\partial_{\mu}\bar{\psi}(x))\gamma^{\mu}-im]=0$ and $[\gamma^{\mu}\partial_{\mu}+im]\psi(x)=0$.

This relation in momentum space becomes
\begin{eqnarray}
q_{\mu}\bar{\psi}(p+q)\gamma^{\mu}\gamma_{5}\psi(p)=2mi\bar{\psi}(p+q)\gamma_{5}\psi(p),
\end{eqnarray}
and this is written for nucleons (the proton $\psi_{p}$ and the neutron $\psi_{n}$ with their masses $m_{N}$ assumed to be degenerate)
\begin{eqnarray}
&&q_{\mu}\bar{\psi}_{p}(p+q)[g_{A}\gamma^{\mu}\gamma_{5}-\frac{2g_{A}m_{N}iq^{\mu}}{q^{2}-\mu^{2}_{\pi}}\gamma_{5}]\psi_{n}(p)|_{\mu^{2}_{\pi}=0} = 0
\end{eqnarray}
where $g_{A}$, the axial current coupling constant of the neutron decay, is multiplied for convenience. The mass $\mu^{2}_{\pi}$ stands for the pion mass which is taken to be $0$ in the present chiral symmetric limit.
The second term in this expression implies that the neutron emits a massless pion which propagates and then decays to $e+\nu$ in the neutron $\beta$ decay. In this picture with emphasis on the virtual pion, one has 
\begin{eqnarray}
g_{A}m_{N}=g_{NN\pi}f_{\pi}
\end{eqnarray}
where $g_{NN\pi}$ stands for the pion-necleon coupling and $f_{\pi}$ stands for the pion decay constant (in the present convention). This is the Goldberger-Treiman relation. Nambu recognized this relation as
an evidence of the spontaneous breaking of chiral symmetry which generated the nucleon mass and the resulting massless Nambu-Goldstone pion. Namely the physical matrix element between the observed neutron and the proton is given by 
\begin{eqnarray}
\int d^{4}e^{-iqx}\langle p(p+q)|j^{\mu}_{5}(x)|n(p)\rangle=\bar{\psi}_{p}(p+q)[g_{A}\gamma^{\mu}\gamma_{5}-\frac{2g_{A}m_{N}iq^{\mu}}{q^{2}}\gamma_{5}]\psi_{n}(p)
\end{eqnarray}
but the axial current for the massless fundamental "nucleon" is conserved
\begin{eqnarray}
\partial_{\mu}j^{\mu}_{5}(x)=0.
\end{eqnarray}
He later emphasized~\cite{nambu-jona-lasinio} {\em
"In order for a chiral invariant Hamiltonian
to allow massive nucleon states and a non-vanishing $j^{\mu}_{5}$
current for q=o, it is therefore necessary to have at the
same time pseudoscalar zero mass mesons coupled with the
nucleons".}

A deep understanding of the spontaneous symmetry breaking in BCS theory thus led to a generalization of the idea to particle physics and to the physical {\em vacuum}, which are expressed in terms of the precise Ward-Takahashi identity.
The idea of spontaneous symmetry breaking is now generalized to a wider domain in physics beyond the superconductivity.

\section{Nambu-Jona-Lasinio model I}
Nambu and Jona-Lasinio considered the famous Lagrangean~\cite{nambu-jona-lasinio}
\begin{eqnarray}
{\cal L}&=&\bar{\psi}(x)i\gamma^{\mu}\partial_{\mu}\psi(x)-g[\bar{\psi}(x)i\gamma_{5}\psi(x)P(x)+\bar{\psi}(x)\psi(x)S(x)]-\frac{1}{2}[P^{2}(x)+S^{2}(x)]
\end{eqnarray}
where we use the auxiliary fields $P(x)$ and $S(x)$ to simplify the following discussion. The original Lagrangean is obtained if one uses the solutions of the equations of motion
\begin{eqnarray}
S(x)=-g\bar{\psi}(x)\psi(x), \ \ \ P(x)=-g\bar{\psi}(x)i\gamma_{5}\psi(x).
\end{eqnarray}
The chiral symmetry is defined by 
\begin{eqnarray}
&&\psi(x)\rightarrow e^{i\alpha \gamma_{5}}\psi(x),\ \ \ \ \bar{\psi}(x)\rightarrow \bar{\psi}(x)e^{i\alpha \gamma_{5}},\nonumber\\
&&S(x)\rightarrow S(x)\cos2\alpha +P(x)\sin2\alpha \nonumber\\
&&P(x)\rightarrow P(x)\cos2\alpha -S(x)\sin2\alpha
\end{eqnarray}
for a global parameter $\alpha$.
The Ward-Takahashi identity is given by (see (A.15))
\begin{eqnarray}
\partial^{x}_{\mu}\langle T^{\star}\bar{\psi}(x)\gamma^{\mu}\gamma_{5}\psi(x)P(y)\rangle=-2i\langle S(x)\rangle\delta^{4}(x-y).
\end{eqnarray}
We now assume the development of the vacuum value $v$ (the order parameter of chiral symmetry)
\begin{eqnarray}
\langle S(x)\rangle=v\neq 0.
\end{eqnarray}
The Ward-Takahashi identity then becomes 
\begin{eqnarray}
\lim_{p\rightarrow 0}
\int dxe^{-ipx}\partial_{\mu}\langle T^{\star}\bar{\psi}(x)\gamma^{\mu}\gamma_{5}\psi(x)P(0)\rangle=-2iv
\end{eqnarray}
which implies the presence of a massless pole in the correlation function
\begin{eqnarray}
\int dxe^{-ipx}\langle T^{\star}\bar{\psi}(x)\gamma^{\mu}\gamma_{5}\psi(x)P(0)\rangle\sim 2v\frac{p^{\mu}}{p^{2}}
\end{eqnarray}
namely, the correlation
\begin{eqnarray}
\langle T^{\star}P(x)P(y)\rangle,
\end{eqnarray}
contains a massless Nambu-Goldstone boson.
This is the standard statement of the {\em Nambu-Goldstone theorem}.

To check this relation, we use the result of the equation of motion for 
$S(x)$
\begin{eqnarray}
v=\langle S(x)\rangle=-g\langle \bar{\psi}(x)\psi(x)\rangle.
\end{eqnarray}
Nambu and Jona-Lasinio evaluated the last term in the lowest one-loop Feynman diagram
\begin{eqnarray}
v&=&g\int \frac{d^{4}p}{(2\pi)^{4}}{\rm Tr}\frac{i}{\pslash -gv}\nonumber\\
&=&4g\int \frac{d^{4}p}{(2\pi)^{4}}\frac{m}{p^{2} +m^{2}}
\end{eqnarray}
where the last expression is written in Euclidean metric, $p^{2}\geq 0$.
By assuming $m=gv \neq 0$, they obtain the self-consistency relation
\begin{eqnarray}
\frac{1}{4g^{2}}=\int \frac{d^{4}p}{(2\pi)^{4}}\frac{1}{p^{2} +m^{2}}.
\end{eqnarray}
They confirm the existence of the solution to this self-consistency condition for a coupling constant 
\begin{eqnarray}
0<\frac{4\pi^{2}}{g^{2}\Lambda^{2}}<1
\end{eqnarray}
when the right-hand side of (4.11) is evaluated by applying a cut-off $\Lambda$ in the momentum space. 

We next start with the lowest order expression of the pseudo-scalar freedom  
$P(x)$
\begin{eqnarray}
\int d^{4}x e^{ipx}\langle T^{\star}P(x)P(0)\rangle=-i
\end{eqnarray}
and evaluate the propagator which includes self-energy corrections, 
\begin{eqnarray}
(-i)[1-i\Sigma+(-i\Sigma)^{2}+.....]=(-i)\frac{1}{1+i\Sigma}.
\end{eqnarray}
In this expression, the one-loop self-energy correction, which consists of a massive fermion loop diagram, is given by 
\begin{eqnarray}
\Sigma(k^{2})
&=&(ig)^{2}\int \frac{d^{4}p}{(2\pi)^{4}}{\rm Tr}\gamma_{5}\frac{i}{\pslash-m}\gamma_{5}\frac{i}{\pslash+\kslash-m}\nonumber\\
&=&4ig^{2}\int_{0}^{1} d\alpha\int \frac{d^{4}p}{(2\pi)^{4}}\frac{m^{2}+p(p+k)}{[p^{2}(1-\alpha)+m^{2}+\alpha(p+k)^{2}+m^{2}]^{2}}\nonumber\\
&=&4ig^{2}\int_{0}^{1} d\alpha\int \frac{d^{4}p}{(2\pi)^{4}}\frac{m^{2}+p^{2}-\alpha(1-\alpha)k^{2}}{[p^{2}+k^{2}\alpha(1-\alpha)+m^{2}]^{2}}.
\end{eqnarray}
We thus obtain the corrected propagator
\begin{eqnarray}
\int d^{4}x e^{ipx}\langle T^{\star}P(x)P(0)\rangle=(-i)\frac{1}{1+i\Sigma(k^{2})}.
\end{eqnarray}
If one uses the self-consistency condition (4.11) in the denominator in this expression, one obtains
\begin{eqnarray}
1+i\Sigma(0)&=&1-4g^{2}\int \frac{d^{4}p}{(2\pi)^{4}}\frac{1}{p^{2}+m^{2}}
\nonumber\\
&=&0,
\end{eqnarray}
namely, the denominator of the propagator of the pseudo-scalar freedom vanishes for $k^{2}=0$. Near $k^{2}\sim 0$, one thus has
\begin{eqnarray}
(-i)\frac{1}{1+i\Sigma(k^{2})}\sim \frac{i}{k^{2}}
\end{eqnarray}
which shows the presence of a {\em massless} boson.

As for the application of the simple ultraviolet cut-off and the lowest order perturbation theory in the above analysis, the authors mention~\cite{nambu-jona-lasinio}: In this connection, it must be kept in mind that our
solutions are only approximate ones. We are operating
under the assumption that the corrections to them are
not catastrophic, and can be appropriately calculated
when necessary. If this does not turn out to be so for
some solution, such a solution must be discarded. 

As it turned out,  the basic reason why this simple one-loop calculation gives a precise result is related to the fact that their calculation is exact in the leading order of large $N$ expansion where 
$N$ stands for a hidden internal freedom of fermions $\psi(x)$, namely, 
if one assumes the existence of many $N$ copies of the fermion.

Nambu and Jona-Lasinio have also shown that\\
(i)The vacuum with spontaneously broken chiral symmetry is orthogonal to the original naive vacuum with $v=0$,\\
(ii)The existence of infinitely many degenerate vacua,\\
(iii) A possible scalar bound state with spin-parity $0^{+}$ and mass $M=2m$.\\

These two features (i) and (ii) characterize the fundamental  aspects of spontaneous symmetry breaking. The last point (iii) played an important role in the later works of Nambu.

\section{Nambu-Jona-Lasinio model II}
In the second paper of Nambu-Jona-Lasinio~\cite{nambu-jona-lasinio2}
, more realistic models of hadrons were discussed. What I found interesting is the model on the doublet structure of nucleons 
\begin{eqnarray}
\psi(x)=\left(\begin{array}{c}
            p(x)\\
            n(x)
            \end{array}\right)
\end{eqnarray}
and the axial isospin current 
\begin{eqnarray}
\vec{j}_{5}^{\mu}(x)=\bar{\psi}(x)\vec{\tau}\gamma^{\mu}\gamma_{5}\psi(x),
\end{eqnarray}
which is originally conserved
\begin{eqnarray}
\partial_{\mu}\vec{j}_{5}^{\mu}(x)=0,
\end{eqnarray}
induces after spontaneous chiral symmetry breaking an iso-triplet of massless $\pi$ mesons
\begin{eqnarray}
\vec{\pi}(x)\vec{\tau}=\pi_{1}(x)\tau_{1}+\pi_{2}(x)\tau_{2}+\pi_{3}(x)\tau_{3}
\end{eqnarray}
together with a iso-singlet $\sigma$ meson with mass
\begin{eqnarray}
m_{\sigma}\simeq 2m
\end{eqnarray}
with $m$ standing for the nucleon mass after the spontaneous symmetry breaking. 

This structure is similar to the linear $\sigma$ model~\cite{gell-mann-levy}, but a very specific mass value. It is my impression that Nambu took  this phenomenon of the massless pseudo-scalar bosons and a very massive scalar boson as a fundamental prediction of the superconductor model of elementary particles. For example, it appears that Nambu tried to
understand the very massive Higgs particle in this picture. At least, the mass of the  observed Higgs mass at 125 GeV is just as heavy as the mass of the top quark mass.

\section{After Nambu-Jona-Lasinio and Nambu-Goldstone theorem}
The Nambu-Jona-Lasinio model (to be precise, its preliminary version) motivated 
Goldstone to write the well-known model for a complex scalar field (he considered several other models also)~\cite{goldstone},  
\begin{eqnarray}
{\cal L}=\partial_{\mu}\phi(x)\partial^{\mu}\phi^{\star}(x)-\lambda(|\phi(x)|^{2}-\frac{1}{2}v^{2})^{2},
\end{eqnarray}
which is invariant under the continuous symmetry $\phi(x)\rightarrow e^{i\beta}\phi(x)$. The spontaneous symmetry breaking implies
\begin{eqnarray}
\langle \alpha| \phi(x)|\alpha\rangle=\frac{v}{\sqrt{2}} e^{i\alpha}
\end{eqnarray}
where we explicitly write the degenerate vacua $|\alpha\rangle$ 
parameterized by $\alpha$. The Lagrangean then becomes
\begin{eqnarray}
{\cal L}&=&\frac{1}{2}\partial_{\mu}\varphi(x)\partial^{\mu}\varphi(x)-\frac{1}{2}
(2\lambda v^{2})\varphi^{2}(x)+\frac{1}{2}\partial_{\mu}\eta(x)\partial^{\mu}\eta(x)\nonumber\\
&&-\frac{\lambda}{4}[\varphi^{4}(x)+4v\varphi^{3}(x)+2\varphi^{2}(x)\eta^{2}(x)+4v\varphi(x)\eta^{2}(x)]
\end{eqnarray}
by parameterising 
\begin{eqnarray}
\phi(x)=\left(\varphi(x)+v+i\eta(x)\right)e^{i\alpha}/\sqrt{2}
\end{eqnarray}
in terms of two real fields $\varphi(x)$ and $\eta(x)$, and the freedom $\eta(x)$ becomes massless.

The field $\phi(x)$ corresponds to 
\begin{eqnarray}
\phi(x)=\frac{1}{\sqrt{2}}[S(x)+iP(x)]
\end{eqnarray}
in the context of the Nambu-Jona-Lasinio model. The appearance of the vacuum value of $S(x)$ implies the massless mode in $P(x)$.

The paper by Goldstone further clarified the nature of spontaneous symmetry breaking
and the generality of the appearance of massless bosons. This paper motivated the formulation of the so-called "Goldstone theorem". The essence of the general proof of this theorem~\cite{goldstone2} is the exact Ward-Takahashi identity (or just called Ward identity at that time) arising from the basic symmetry in the starting Lagrangean. In this aspect, Nambu extensively used the Ward-Takahashi identity in his reformulation of BCS theory and Nambu and Jona-Lasinio
emphasized the Ward-Takahashi identity in their formulation of the appearance 
of the massless "pion". In contrast, Goldstone did not mention the Ward-Takahashi
identity in his paper~\cite{goldstone}. Goldstone emphasized the basic mechanism of spontaneous symmetry breaking in the framework of renormalizable field theory and the general appearance of massless bosons, while Nambu used the Ward-Takahashi identity as a basic means to express the idea of Noether's theorem and asserted the dynamical appearance of massless bosons. In this sense, it is a fair custom to call the general appearance of massless bosons in the spontaneous breaking of continuous (bosonic) symmetry  as {\em Nambu-Goldstone theorem} and the resulting bosons as {\em Nambu-Goldstone bosons}.   

The model (6.1) or closely related models were later used in the formulation of the Higgs mechanism and the Higgs boson~\cite{higgs}. In the decomposition (6.4), the freedom $\eta(x)$ in the direction orthogonal to the vacuum value is the (would-be) Nambu-Goldstone boson, and the freedom $\varphi(x)$ in the radial direction $v$ corresponds to the particle later identified as the {\em Higgs boson}~\cite{higgs}.
\\

Nambu together with Lurie~\cite{nambu-lurie} also formulated the idea of the soft pion theorem, a low energy theorem which dictates the soft-pion emission
from the nucleon in collision phenomena, which is analogous to the photon emission from a decelerated charged particle.\\

As for the interpretation of the Goldberger-Treiman relation, Nambu's picture 
competed with the idea of the PCAC (partially conserved axial-vector current)
hypothesis~\cite{gell-mann-levy}
\begin{eqnarray}
\partial_{\mu}j^{\mu}_{5}(x)=-c\pi(x)
\end{eqnarray}
with a constant $c$, which states the idea that the Goldberger-Treiman relation holds since the mass of the pion happens to be very small compared to other hadrons; this picture itself was also mentioned by Nambu in his original paper~\cite{nambu-chiral}. Namely, the competition with the opinion, which prefers the simple relation (6.6) and asserts that the spontaneous symmetry breaking and the resulting massless pion, as Nambu's idea implies, are not inevitable, continued till around 1970.

It is my impression that only after the experimental establishment of the Standard Model and the Higgs mechanism, the deep idea of the physical vacua with spontaneously broken symmetry of Nambu was universally appreciated.

\section{Han-Nambu model}
Han and Nambu~\cite{han-nambu} started their paper by noting the problems to be resolved in the original idea of quarks by Gell-Mann:
Although the $SU(6)$ symmetry strongly indicates
that the baryon is essentially a three-body
system built from some basic triplet field or fields, the
quark model~\cite{gell-mann} is not entirely satisfactory from a realistic
point of view, because\\
(a) the electric charges are not integral,\\
(b) three quarks in $s$ states do not form
the symmetric $SU(6)$ representation assigned to the
baryons, and\\
(c) a simple dynamical mechanism is lacking
for realizing only zero-triality states as the low-lying
levels.
 
The original quark model of Gell-Mann deals with a single triplet
\begin{eqnarray}
\left(\begin{array}{c}
            u(\frac{2}{3})\\
            d(-\frac{1}{3})\\
            s(-\frac{1}{3})
            \end{array}\right)
\end{eqnarray} 
while the Sakata triplet~\cite{sakata} is 
\begin{eqnarray}
\left(\begin{array}{c}
            p(+1)\\
            n(0)\\
            \Lambda(0)
            \end{array}\right)
\end{eqnarray}
where the charge assignment of the quark (and Sakata triplet) is written explicitly.

As Han and Nambu mentioned, the success of $SU(6)$ classification of baryons~\cite{gursey} implies that 3 quarks need to be put into the symmetric $S$ state violating the exclusion principle of Pauli if one understands quarks as  fermions. 
Han and Nambu thus proposed to extend the original quark model to a 3-triplet model and classify the quarks by $SU(3)\otimes SU^{\prime}(3)$, where
$SU^{\prime}(3)$ is a new symmetry introduced together with 3 varieties of quarks. They then suggest the possible integral charge assignment of quarks as 
\begin{eqnarray}
\left(\begin{array}{ccc}
            u(1)&u(0)&u(1) \\
            d(0)&d(-1)&d(0)\\
            s(0)&s(-1)&s(0)
            \end{array}\right).
\end{eqnarray} 
This is a very clever idea. By considering the singlet states with respect to
the new $SU^{\prime}(3)$, which is related to the binding mechanism of quarks, as lowest lying states,  one can explain the success of quark model at low energies by avoiding the difficulty related to the Pauli principle.  In general, the masses of hadrons will then depend on the Casimir operators
of $SU(3)^{\prime}$. For example, a simple linear form will be
\begin{eqnarray}
m= m_{0}+ m_{2}C_{2}^{\prime}
\end{eqnarray}
where $C_{2}^{\prime}$ is the eigenvalue of quadratic Casimir operator of 
$SU(3)^{\prime}$. Since $C_{2}^{\prime}$ increases with the dimensionality of
representation, the lowest mass levels will be $SU(3)^{\prime}$ singlets. The new $SU^{\prime}(3)$ symmetry, which is active to define the charges of quarks, is {\em visible} in this scheme.

This scheme predicts the so-called $R$  ratio in $e^{+}e^{-}$ annihilation
to be~\cite{nambu-han},
\begin{eqnarray}
R=\sum_{i}Q^{2}_{i}=4
\end{eqnarray}
(with $i$ running over all the possible quarks) at low energies below the charm threshold, while the modern QCD~\cite{gell-mann2}
\begin{eqnarray}
\left(\begin{array}{ccc}
            u(\frac{2}{3})&u(\frac{2}{3})&u(\frac{2}{3}) \\
            d(-\frac{1}{3})&d(-\frac{1}{3})&d(-\frac{1}{3})\\
            s(-\frac{1}{3})&s(-\frac{1}{3})&s(-\frac{1}{3})
            \end{array}\right).
\end{eqnarray} 
predicts $R=\sum_{i}Q^{2}_{i}=2$. 

As for the pion decay $\pi_{0}\rightarrow \gamma\gamma$~\cite{jackiw}, the decay
amplitude has a structure (see also (A.17))
\begin{eqnarray}
A(\pi^{0}\rightarrow \gamma\gamma)\propto S\epsilon^{\mu\nu\alpha\beta}F_{\mu\nu}F_{\alpha\beta}
\end{eqnarray}
where $\epsilon^{\mu\nu\alpha\beta}$ is the antisymmetric Levi Civita symbol and $F_{\mu\nu}$ is the Maxwell field strength tensor. The parameter $S$ is related to 
\begin{eqnarray}
\frac{1}{2}[Q^{2}-(Q-1)^{2}]=Q-\frac{1}{2}
\end{eqnarray}
for {\em each} quark triplet with charges $(Q,Q-1,Q-1)$. Both of Han-Nambu and QCD predict the correct value; to be explicit, $S=3(\frac{2}{3}-\frac{1}{2})=\frac{1}{2}$ for QCD, $S=2(1-\frac{1}{2})+(-\frac{1}{2})=\frac{1}{2}$ for Han-Nambu, and $S=1-\frac{1}{2}=\frac{1}{2}$ for the Sakata model. Han and Nambu~\cite{nambu-han} also emphasized the equivalence of QCD-type charge assignment with the parafermi statistics of order 3 suggested by Greenberg~\cite{greenberg}; of course, QCD contains the massless $SU(3)$ Yang-Mills field as the force field and in this respect quite different. In 1974 before the discovery of $J/\psi$ (charm quark), the Han-Nambu scheme appeared to be favored~\cite{nambu-han}, but after the discovery of the charm (and also the heavy lepton $\tau$) and the spectroscopy of related mesonic states eventually disfavored the Han-Nambu scheme. Yet the original idea of Han and Nambu, which motivated to resolve the problems listed at the beginning of this section, is highly appreciated. 

When I looked at the papers of Han and Nambu~\cite{han-nambu, nambu-han} at this time, I recognized that their papers contain no expressions of the Lagrangean
such as the Yang-Mills action~\cite{yang}. I always thought that they were the first to write the Lagrangean which is called QCD nowadays, but to my surprise, this was not the case. One of the reasons for this absence of the Lagrangean may be that field theory for the strong interaction was very unpopular in the middle of 1960s.

\section{Discussion and conclusion}
Nambu lived in a revolutionary era of fundamental physics, and he contributed to make the revolution more dramatic.
In 1957, the long standing problem of the construction of a microscopic theory of 
superconductivity was solved by BCS. The solution was such that a charged electron pair condense in the ground state to make it more stable than the naive vacuum state. This picture alerted Nambu in connection with gauge invariance. He thus studied this theory very carefully. Nambu later mentioned privately that this was a kind of side business for him
since he was a particle theorist and busy to write papers on particle physics. But this study led to an enormously rich idea of spontaneous symmetry breaking in general.
On the other hand, a surprising breaking of parity symmetry in weak interactions was suggested by Lee and Yang in 1956. This led to a recognition of the  importance of $\gamma_{5}$ and related chiral symmetry in Dirac theory. The Fermi theory of weak interactions was now completed 
to be a universal $V-A$ theory of current-current interactions. Based on the experience of the spontaneous symmetry breaking, Nambu immediately recognized that the spontaneous symmetry breaking of chiral symmetry is realized in nature, although in an approximate manner, as the Goldberger-Treiman relation.  

He thus constructed the celebrated Nambu-Jona-Lasinio model of elementary particles in analogy with superconductivity. This generalized notion of spontaneous symmetry breaking was further clarified by Goldstone, and soon after, the Nambu-Goldstone theorem was established. The pions were recognized to be Nambu-Goldstone bosons and in fact pions may be said to be the only known (clear-cut) Nambu-Goldstone bosons in particle physics so far. This idea of spontaneous symmetry breaking in our vacuum led to the recognition of the Higgs mechanism, and eventually, to the establishment of the Standard Model with the Weinberg-Salam scheme incorporated~\cite{weinberg}.

In the forefront of strong interactions, the quark model of Gell-Mann was proposed in 1964 with a revolutionary idea of fractional electric charges. This time, Nambu tried to maintain the more conventional idea of integrally charged quarks, and thus proposed a Han-Nambu version of the quark model with extra non-Abelian freedom. The gluon at that time was not known to be something similar to massive vector mesons in the manner of Sakurai~\cite{sakurai} or more fundamental ones analogous to Maxwell field. This issue was eventually settled to be massless pure Yang-Mills field as the gluon, and QCD with fractionally charged quarks as we know nowadays was established. 

Nambu contributed greatly to the construction of the basic building blocks of those fundamental developments in the years of 1960-1965.
\\
 
\section*{Acknowledgments}

 I would like to express my deep gratitude to Professor Yoichiro  Nambu, who
 encouraged me at various occasions in my research starting my postdoc days at Chicago in 1970-1972.

\appendix
 
\section{Path integral formulation of Noether's theorem}

The Noether's theorem states that any continuous symmetry in the action defined in terms of the Lagrangean
implies the existence of a conserved quantity. The Ward-Takahashi  identity
in field theory is a manifestation of this theorem in terms of Green's functions.  

We illustrate the above idea in field theory using QED as an example, which describes the Dirac field  $\psi(t,\vec{x})$ (the electron) and the gauge field $A_{\mu}(t,\vec{x})$ (the electromagnetic field).
The path integral is defined by applying a suitable gauge fixing,
\begin{eqnarray}
\int{\cal D}\psi{\cal D}\bar{\psi}{\cal D}A_{\mu} \exp\{\frac{i}{\hbar}\int d^{4}x{\cal L}(\bar{\psi},\psi,A_{\mu})\}
\end{eqnarray}
in terms of the Lagrangean
\begin{eqnarray}
{\cal L}(\bar{\psi},\psi,A_{\mu})
&=&\bar{\psi}i\gamma^{\mu}[\partial_{\mu}-ieA_{\mu}]\psi-m\bar{\psi}\psi-\frac{1}{4}(\partial_{\mu}A_{\nu}-\partial_{\nu}A_{\mu})^{2}.
\end{eqnarray}

\noindent{\bf Current conservation:}\\

The above Lagrangean (to be precise the action) is invariant under the global phase transformation
$\psi\rightarrow e^{i\epsilon}\psi,\ \ \ \bar{\psi}\rightarrow \bar{\psi}e^{-i\epsilon}$,
for a constant $\epsilon$. If one localizes the symmetry transformation $\epsilon\rightarrow \epsilon(x)$,
\begin{eqnarray}
\psi^{\prime}(x)= e^{i\epsilon(x)}\psi(x),\ \ \ \bar{\psi}^{\prime}(x)=\bar{\psi}e^{-i\epsilon(x)},
\end{eqnarray}
one obtains the identity
\begin{eqnarray}
&&\int{\cal D}\psi{\cal D}\bar{\psi}{\cal D}A_{\mu} \exp\{\frac{i}{\hbar}\int d^{4}x{\cal L}(\bar{\psi},\psi,A_{\mu})\}\nonumber\\
&&=\int{\cal D}\psi^{\prime}{\cal D}\bar{\psi}^{\prime}{\cal D}A_{\mu} \exp\{\frac{i}{\hbar}\int d^{4}x{\cal L}(\bar{\psi}^{\prime},\psi^{\prime},A_{\mu})\},
\end{eqnarray}
namely, the value of the integral is independent of the naming of the integration variable. We have
\begin{eqnarray}
{\cal L}(\bar{\psi}^{\prime},\psi^{\prime},A_{\mu})&=&{\cal L}(\bar{\psi},\psi,A_{\mu})-\frac{i}{\hbar}\int d^{4}x\partial_{\mu}\epsilon(x)\bar{\psi}(x)\gamma^{\mu}\psi(x).
\end{eqnarray}
We furthermore assume the invariance of the path integral measure
\begin{eqnarray}
{\cal D}\psi^{\prime}{\cal D}\bar{\psi}^{\prime}={\cal D}\psi{\cal D}\bar{\psi}
\end{eqnarray}
under the transformation, we then obtain the identity
\begin{eqnarray}
&&\int{\cal D}\psi{\cal D}\bar{\psi}{\cal D}A_{\mu}\left(-\frac{i}{\hbar}\int d^{4}x\partial_{\mu}\epsilon(x)\bar{\psi}(x)\gamma^{\mu}\psi(x)\right)\exp\{\frac{i}{\hbar}\int d^{4}x{\cal L}\}\nonumber\\
&&\equiv\langle\frac{i}{\hbar}\int d^{4}x\epsilon(x)\partial_{\mu}(\bar{\psi}(x)\gamma^{\mu}\psi(x))\rangle=0,
\end{eqnarray}
namely, we conclude the current conservation condition
\begin{eqnarray}
\partial_{\mu}\langle\bar{\psi}(x)\gamma^{\mu}\psi(x)\rangle=0,
\end{eqnarray}
since $\epsilon(x)$ is arbitrary.
\\

\noindent{\bf Chiral identity:}\\

The Lagrangean of QED, except for the mass term, is invariant under the global chiral ($\gamma_{5}$) transformation
$\psi\rightarrow e^{i\epsilon\gamma_{5}}\psi,\ \ \ \bar{\psi}\rightarrow \bar{\psi}e^{i\epsilon\gamma_{5}}$.
We thus consider the localized transformation
\begin{eqnarray}
\psi^{\prime}(x)= e^{i\epsilon(x)\gamma_{5}}\psi(x),\ \ \ \bar{\psi}^{\prime}(x)=\bar{\psi}e^{i\epsilon(x)\gamma_{5}}
\end{eqnarray}
and obtain the identity
\begin{eqnarray}
&&\int{\cal D}\psi{\cal D}\bar{\psi}{\cal D}A_{\mu} \exp\{\frac{i}{\hbar}\int d^{4}x{\cal L}(\bar{\psi},\psi,A_{\mu})\}\nonumber\\
&&=\int{\cal D}\psi^{\prime}{\cal D}\bar{\psi}^{\prime}{\cal D}A_{\mu} \exp\{\frac{i}{\hbar}\int d^{4}x{\cal L}(\bar{\psi}^{\prime},\psi^{\prime},A_{\mu})\},
\end{eqnarray}
namely, the change of naming of integration variables does not change the integral itself. 
If one recalls 
\begin{eqnarray}
&&\frac{i}{\hbar}\int d^{4}x{\cal L}(\bar{\psi}^{\prime},\psi^{\prime},A_{\mu})\nonumber\\
&&=\frac{i}{\hbar}\int d^{4}x{\cal L}(\bar{\psi},\psi,A_{\mu})
+\frac{i}{\hbar}\int d^{4}x\epsilon(x)[\partial_{\mu}\bar{\psi}(x)\gamma^{\mu}\gamma_{5}\psi(x)-2mi\bar{\psi}(x)\gamma_{5}\psi(x)],
\end{eqnarray}
and assumes the invariance of the path integral measure
\begin{eqnarray}
{\cal D}\psi^{\prime}{\cal D}\bar{\psi}^{\prime}={\cal D}\psi{\cal D}\bar{\psi},
\end{eqnarray}
one obtains the identity
\begin{eqnarray}
&&\int{\cal D}\psi{\cal D}\bar{\psi}{\cal D}A_{\mu}\{\frac{i}{\hbar}\int d^{4}x[\epsilon(x)\partial_{\mu}\bar{\psi}(x)\gamma^{\mu}\gamma_{5}\psi(x)-2mi\bar{\psi}(x)\gamma_{5}\psi(x)]\}\exp\{\frac{i}{\hbar}\int d^{4}x{\cal L}\}\nonumber\\
&&=\langle\frac{i}{\hbar}\int d^{4}x\epsilon(x)[\partial_{\mu}(\bar{\psi}(x)\gamma^{\mu}\gamma_{5}\psi(x))-2mi\bar{\psi}(x)\gamma_{5}\psi(x)]\rangle=0
\end{eqnarray}
namely, the {\em partial} conservation condition of axial (or chiral) current
\begin{eqnarray}
\partial_{\mu}\langle\bar{\psi}(x)\gamma^{\mu}\gamma_{5}\psi(x)\rangle=\langle2mi\bar{\psi}(x)\gamma_{5}\psi(x)\rangle.
\end{eqnarray}
Because of the breaking of the chiral symmetry by the mass term, the current is not completely conserved. If one starts with $\int{\cal D}\psi{\cal D}\bar{\psi}{\cal D}A_{\mu}\bar{\psi}(y)i\gamma_{5}\psi(y) \exp\{\frac{i}{\hbar}\int d^{4}x{\cal L}(\bar{\psi},\psi,A_{\mu})\}$, one obtains the identity
\begin{eqnarray}
\partial^{x}_{\mu}\langle\bar{\psi}(x)\gamma^{\mu}\gamma_{5}\psi(x)\bar{\psi}(y)i\gamma_{5}\psi(y)\rangle=-2i\langle\bar{\psi}(y)\psi(y)\rangle\delta^{4}(x-y)
\end{eqnarray}
in the case of the massless fermion. This relation is used in (4.4).
\\

\noindent {\bf Chiral anomaly}~\cite{jackiw}:
\\

To be precise, in the present $U(1)$ type chiral transformation the path integral measure is not invariant and thus the naive Neother's theorem is not valid. One obtains the non-trivial Jacobian factor for the chiral transformation~\cite{fujikawa} 
\begin{eqnarray}
{\cal D}\psi^{\prime}{\cal D}\bar{\psi}^{\prime}={\cal D}\psi{\cal D}\bar{\psi}\exp\{ i\int d^{4}x \epsilon(x)\frac{e^{2}}{16\pi^{2}}\epsilon^{\mu\nu\alpha\beta}F_{\mu\nu}F_{\alpha\beta}\}
\end{eqnarray}
and one obtains the (anomalous) chiral identity
\begin{eqnarray}
\partial_{\mu}\langle\bar{\psi}(x)\gamma^{\mu}\gamma_{5}\psi(x)\rangle&=&\langle2mi\bar{\psi}(x)\gamma_{5}\psi(x)\rangle+\langle\frac{e^{2}}{16\pi^{2}}\epsilon^{\mu\nu\alpha\beta}F_{\mu\nu}F_{\alpha\beta}\rangle,
\end{eqnarray}
where $F_{\mu\nu}$ stands for the Maxwell field strength tensor with $\epsilon^{\mu\nu\alpha\beta}$ the Levi Civita symbol. If one understands that the field $\psi(x)$ is standing for quarks, the above identity is crucial to analyze the neutral pion decay $\pi^{0}\rightarrow \gamma\gamma$~\cite{jackiw}. See eq.(7.7).

In the context of QCD, the last term in the above identity is replaced by Yang-Mills field strength tensor  and it is essential to analyze the spontaneous chiral symmetry breaking and the presence or absence of the  Nambu-Goldstone boson $\eta^{\prime}$. In the case of pions discussed by Nambu and Jona-Lasinio, this QCD type anomalous identity is not important.

\end{document}